\documentclass[12pt]{article}
\newcommand{\beq}[3]{\begin{equation}  \label{#1#2#3}}
\newcommand{\eeq}{ \end{equation}}
\newcommand{\ba}{\begin{array}}
\newcommand{\ea}{\end{array}}

\newcommand{\remark}[1]{}
\let\LARGE=\Large
\let\Large=\large

\newcommand{\be}[3]{\begin{equation}  \label{#1#2#3}}     
\newcommand{\ee}{ \end{equation}}
\newcommand{\bea}{\begin{eqnarray}}
\newcommand{\eea}{\end{eqnarray}}

\newcommand{\ft}[2]{{\textstyle\frac{#1}{#2}}}

\def\beq{\begin{equation}}
\def\eeq{\end{equation}}
\def\beqa{\begin{eqnarray}}
\def\eeqa{\end{eqnarray}}

\begin{document}


\thispagestyle{empty}
\rightline{HU-EP-01/06}
\rightline{hep-th/0102128}

\vspace{15truemm}

\centerline{\bf \LARGE
Curved BPS domain wall solutions in
}
%
\vspace{3mm}
\centerline{\bf \LARGE 
four-dimensional ${\cal N}=2$
supergravity }

\bigskip

\vspace{2truecm}

\centerline{\bf Klaus Behrndt\footnote{e-mail: 
 behrndt@physik.hu-berlin.de} ,
Gabriel Lopes Cardoso\footnote{e-mail:
 gcardoso@physik.hu-berlin.de}
{\rm and}
Dieter L\"ust\footnote{e-mail: luest@physik.hu-berlin.de}
}

\vspace{1truecm}

\centerline{\em  Institut f\"ur Physik, Humboldt University}
\centerline{\em Invalidenstra\ss{}e 110, 10115 Berlin, Germany}

\vspace{2truecm}


\begin{abstract}

We construct four-dimensional domain wall solutions of 
${\cal N}=2$ gauged supergravity coupled to
vector and to hypermultiplets.  The gauged supergravity theories 
that we consider
are obtained by performing two types of Abelian gauging.  In both cases we
find that the behaviour of the scalar fields belonging to the vector
multiplets is governed by the so-called attractor equations
known from the study of BPS
black hole solutions in ungauged
${\cal N}=2$ supergravity theories.
The scalar fields belonging to the 
hypermultiplets, on the other hand, are either constant or exhibit a run-away
behaviour.  These domain wall solutions preserve $1/2$ of supersymmetry
and they 
are, in general, curved.  We briefly comment on the amount of supersymmetry
preserved by domain wall solutions in gauged supergravity theories obtained
by more general gaugings.

\end{abstract}




\newpage


\section{Introduction}


Recently there has been a lot of interest in constructing domain wall solutions
in gauged supergravity. 
As consequence of the AdS/CFT correspondence 
\cite{Maldacena:1998re,Witten:1998qj,Gubser:1998bc}, 
domain wall solutions can
provide the supergravity duals of field theories near a fixed point and may,
in addition, give a holographic picture of the renormalization
group (RG) flow. 
In order to understand this in more detail it is
important to have explicit solutions on the supergravity side.

The BPS domain wall
solutions of five-dimensional gauged supergravity theories
describe the RG flow between 
superconformal field theories in four dimensions 
(for an example see \cite{Freedman:1999gp}).
Here, the gravitational $AdS_5$ backgrounds arise as the near horizon
geometry of type IIB
D3-branes located at
certain transversal six-dimensional spaces.
In the same way, four-dimensional $AdS_4$ supergravities 
emerge, 
for example,
from M-theory membranes located at certain transversal 
eight-dimensional spaces.  These
are then holographically
related to
three-dimensional conformal field theories \cite{Fre}, 
which are however not very
well understood at the present time. 
In this paper we will focus on BPS domain wall solutions of 
gauged supergravity theories with eight supercharges (gauged ${\cal N}=2$
supergravity) in four dimensions. 
Domain wall solutions in theories with less supersymmetries
were previously considered in \cite{Cvetic:1991vp,Cvetic:1992bf}. 

The four-dimensional ${\cal N}=2$ gauged supergravity theories we will 
consider are obtained by performing 
a gauging either of 
a $U(1)$ subgroup of the $SU(2)$ R-symmetry or of a particular
Abelian Killing symmetry  of the universal hypermultiplet moduli space.
The former results in models with a potential term
for the vector scalars only, whereas the latter
results in
a potential term for the vector scalars and for the dilaton
field entering the universal hypermultiplet. 
In the context of string theory, the latter models 
can be obtained by Calabi-Yau threefold
compactifications of type II string
theory in the presence of internal H-fluxes \footnote{It is conceivable 
\cite{Karch:1998yv,culu}
that string Calabi-Yau compactifications with internal H-fluxes in the field
theory limit are T-dual to D3-branes in the presence of
transversal RR-fluxes \cite{Polchinski:2000uf}.}
\cite{020}-\cite{  
160}.
The vacuum structure of these type II Calabi-Yau threefold compactifications
was discussed in some detail in 
\cite{020}-\cite{
160}.
It was shown that ${\cal N}=2$
supersymmetric ground states with {\it flat}
four-dimensional Minkowski spacetime
are only possible in certain degeneration limits of the underlying
Calabi-Yau spaces (otherwise ${\cal N}=2$ supersymmetry gets completely
broken at generic points in the Calabi-Yau moduli space).
Thus, in general, non-vanishing
H-fluxes do allow for supersymmetric anti-de-Sitter
groundstates with non-vanishing cosmological constant $\Lambda_4$.
This implies that the associated four-dimensional field equations
possess solutions describing non-flat gravitational backgrounds, such as
domain walls. 

In this paper we will construct four-dimensional domain wall solutions 
of the gauged supergravity theories described above.
We will show that (as already indicated in \cite{160,190})
the behaviour of the scalar fields belonging to the vector
multiplets is governed by a set of equations 
already
known from the study of BPS
black hole solutions in ungauged
${\cal N}=2$ supergravity theories, 
namely the so-called
attractor equations 
\cite{Fer:1996dd}-\cite{
110}.  
We will also show that 
in general, the four-dimensional BPS equations are solved by
curved walls,
i.e. there is in general also a three-dimensional cosmological constant
$\Lambda_3$ on the domain wall.  
In fact, the
curvature of the wall must cancel contributions coming from an expression
closely related to the 
U(1)-K\"ahler connection (as we will see, this 
effect already occurs in the context of pure ${\cal N}=2$ supergravity).
This is in 
contrast with what happens in
five dimensions in analogous models.  
Since such a connection is 
not present in five dimensions (and whenever the
pullback of the 
$SU(2)$ connection is trivial), the
five-dimensional BPS equations only allow for 
flat domain wall solutions \cite{010, 140}.

We can then give the following
dictionary between the quantities which play a role in black hole
solutions of ungauged ${\cal N}=2$ supergravity and those of the 
corresponding domain
wall solutions of  gauged ${\cal N}=2$ supergravity:

\begin{itemize}

\item 
In  gauged ${\cal N}=2$ supergravity we will use the following symplectic
invariant superpotential:
\be001
W =  
{\rm e}^{s \phi} \, \Big( \alpha_I L^I - \beta^I
M_I \Big) \quad , \quad s=0,2\, ,
\ee 
where the symplectic vector $(L^I,M_I)$ 
($I=0,\dots ,N_V$) denotes the sections which depend on the 
vector scalar fields, and $\phi$ is the dilaton field. 
In the case of the gauging of a particular Abelian isometry of the universal
hypermultiplet moduli space,
the 
entries of the constant
symplectic vector $(\alpha_I,\beta^I)$ correspond to the electric/magnetic 
$U(1)^{N_V+1}$ charges
of the universal hypermultiplet and emerge
in type IIA/B string compactifications as internal H-fluxes on Calabi-Yau
threefolds with $h^{1,1}(h^{2,1})=N_V$ for type IIA (type IIB) \cite{030,040}.
The dependence of this superpotential on the vector scalars is, on the other
hand, the direct analogue of the central charge $Z$ of the ${\cal N}=2$
supersymmetry algebra that plays a role in the context of BPS
black holes in ungauged ${\cal N}=2$ supergravity. 
Here, the $(\alpha_I,\beta^I)$
are just the electric/magnetic $U(1)^{N_V+1}$ charges of the black holes.

\item 
The attractor equations (see eq.(\ref{stab}))
determine the running of the vector scalar fields from the domain wall to the
supersymmetric extrema, $DW=0$, reached at spatial infinity. 
In fact,  the vector scalar fields are stabilized in the sense
that their values at spatial infinity only depend on the constants
$(\alpha_I,\beta^I)$. 
Therefore the attractor equations  are relevant
for the RG-flow in the corresponding field theories.
In the context of ${\cal N}=2$ black holes the same stabilization
equations determine the evolution of the vector scalars from spatial infinity
towards the horizon, where the central charge is extremized, $DZ=0$, and
the scalars
are entirely expressed in terms 
of the $(\alpha_I,\beta^I)$.

\item
For ${\cal N}=2$ domain walls, the extremum of the potential with respect
to the vector scalar fields,
$V_{\rm extr.}\sim
|W|^2$, is again a function only of the $(\alpha_I,\beta^I)$
and corresponds to a 
four-dimensional
cosmological constant $\Lambda_4$ at spatial infinity. The analogous quantity,
$|Z|^2_{\rm extr.}$, is just the entropy ${\cal S}$
of the ${\cal N}=2$ black holes,
i.e. the area of the horizon.

\item
We will show that the quantity (see section 3)
\be002
{\cal A}^Y = \ft{1}{2} {\rm e}^{-2U} (Y^I - {\bar Y}^I) 
\stackrel{\leftrightarrow}
{\rm d}
(F_I - {\bar F}_I )
\ee
provides the three-dimen\-sio\-nal cosmological constant 
$\Lambda_3$ on the domain wall.
This means that a non-vanishing one-form ${\cal A}^Y$ leads to curved
anti-de-Sitter like
domain wall solutions. On the black hole side, the same object
corresponds to the angular momentum of stationary but in general non-static
solutions \cite{130,230,110}.

\item 
Finally,
closely related brane configurations play a role in the comparison of
${\cal N}=2$ domain wall and black hole solutions.
In type IIA superstring compactified on a Calabi-Yau threefold $M$,
D4-branes, being $\beta^A$-times
wrapped around  4-cycles of $M$, together with $\alpha_0$ D0-branes,
lead to black holes with magnetic charges $p^A$ and one electric
charge $\alpha_0$. The corresponding entropy is determined by the triple
intersection form $C_{ABC}$.
Increasing the dimensionality of the branes by two, i.e. considering 
boundstates of  wrapped D6-branes together with $\alpha_0$ D2-branes,
leads to membranes in four dimensions, which represent the source
for the supergravity domain wall solutions in four uncompactified dimensions.
Similarly, in type IIB theory compactified on a Calabi-Yau threefold, D3-branes
wrapped around 3-cycles provide charged black hole solutions, whereas
the same configurations of wrapped
D5-branes correspond to domain wall solutions.

\end{itemize}

The paper is organized as follows.
In the next section we will recall certain aspects of gauged ${\cal N}=2$
supergravity in four dimensions. 
In section 3 we will solve the BPS equations for domain wall
solutions. Section 4 contains a discussion of our solutions
as well as various concluding remarks.


\section{${\cal N}=2$ gauged supergravity
}
\setcounter{equation}{0}

Domain walls are codimension one solutions that separate the spacetime into
regions corresponding to different vacua.  In the simplest case,
a domain wall is 
supported by a gauge potential that couples to its world volume.  The field
strength of this gauge potential is dual to a 
cosmological constant. In a more general setting with non-trivial couplings
to scalar fields, this cosmological
constant appears as an extremum of the potential term in the Lagrangian.  The
resulting solution then describes a 
flow towards an
extremum, and if the potential possesses several extrema, the solution may  
interpolate between them.  In the following we will be 
interested in constructing
domain wall solutions in ${\cal N}=2$ 
gauged supergravity theories in four dimensions.

The ${\cal N}=2$ gauged supergravity theories that we consider are based on 
abelian vector
multiplets (labelled by an index $I = 0, \dots, N_V$) 
and hypermultiplets coupled to the ${\cal N}=2$ supergravity fields. 
Potentials that are allowed by ${\cal N}=2$ supersymmetry
are then obtained by performing a 
gauging of (some of) the various global symmetries. 
There are two different types of gaugings, namely
(i) one can either gauge
some of 
isometries of the moduli space of ungauged ${\cal N}=2$ supergravity or (ii) 
one can gauge (part of) 
the $SU(2)$ R-symmetry, which only acts on the fermions. 
In ${\cal N}=2$
supergravity, at the two-derivative level,
the moduli space is a direct product ${\cal M} = {\cal
M}_{V} \times {\cal M}_{H}$, where ${\cal M}_{V}$ and ${\cal M}_{H}$ are
parameterized by the
scalars belonging to the 
vector multiplets and to the 
hypermultiplets, respectively. 
In the following, we will only consider Abelian gaugings, either of 
(some of) the isometries of the hyper scalar manifold ${\cal M}_H$ 
or of 
the $SU(2)$ $R$-symmetry of ${\cal N}=2$ supergravity.
We refer to \cite{210,b1,b2} for a detailed
description of the gauging.

The hyper scalar manifold ${\cal M}_{H}$ is a quaternionic
space, and hence it possesses three complex structures $J^x$ as well as a
triplet of K\"ahler two-forms $K^x$ (here $x =1,2,3$
denotes an $SU(2)$ index). 
The holonomy group is
$SU(2) \times Sp(n_H)$ and the K\"ahler forms have to be covariantly
constant with respect to the $SU(2)$ connection. The isometries 
of ${\cal M}_{H}$ are
generated by a set of Killing vectors $k_I^u$,
\be090
q^u \rightarrow q^u + k^u_I \epsilon^I \;,
\ee
and therefore the gauging of (some of) the Abelian isometries results in the
introduction of 
gauge covariant derivatives via the replacement 
$dq^u \rightarrow
d q^u + k_I^u A^I$. In order to maintain supersymmetry, the gauging 
of the isometries has to
preserve the quaternionic structure, which implies that the Killing
vectors have to be tri-holomorphic. 
This is the case whenever it is possible to 
express the Killing vectors
in terms of a triplet of real Killing prepotentials ${\cal P}^x_I$,
as follows:
\be080
K^x_{uv} k^v_I = - \nabla_u {\cal P}_I^x \equiv - \partial_u {\cal P}_I^x -
\epsilon^{xyz} \omega^y_u {\cal P}^z_I \ .
\ee
Here $\omega_u^y$ are the $SU(2)$ connections, which are related 
to the K\"ahler
forms by $K^x_{uv} = - \nabla_{[u} \omega^x_{v]}$. 
By using the Pauli matrices
$\sigma^x$ one can also revert to matrix notation and write 
\be099
({\cal P}_I)_{ij} = \sum_{x=1}^3  
{\cal P}^x_I \,  (\sigma^x)_i^{\ k} \, 
\varepsilon_{jk} \;\;,
\ee
where $\varepsilon_{jk}$ denotes the two-dimensional 
antisymmetric $\varepsilon$-symbol.

In this paper we
will, for concreteness, only consider the quaternionic space
associated with the so-called universal hypermultiplet.
Classically, it is given by the coset space $SU(2,1)/U(2)$.
Its 
K\"ahler potential is, in a certain parameterization, given by
\be140 
K = - \log [S+\bar S - 2(C + \bar C)^2] \;.
\ee 
This coset space has two Abelian isometries which are generated
by the Killing vectors 
associated to shifts in the imaginary parts of $S$ and $C$.
The gauging of these two Abelian isometries has been discussed in \cite{030}
(we refer to
\cite{010,ceresole,260,270} for a discussion of 
the gauging of (some of) the other isometries of 
the universal hypermultiplet).
In this paper, however, we will only consider the case 
when the shift in the imaginary part of $S$ is gauged.  The associated
Killing vector is given by \cite{030}  
\be150
k_I = k_I^u \frac{\partial}{\partial q^u} = - \alpha_I \,  
i(\partial_S - \partial_{\bar S})  \;,
\ee
where the $\alpha_I$ are constant and real.
This is the four-dimensional 
analogue of the gauging of five-dimensional supergravity discussed in
\cite{010}.  The case where one independently gauges both the Abelian
isometries results
in a potential which makes the construction of explicit domain wall
solutions 
rather difficult, and will not be discussed here.

The Killing prepotential associated with the Killing vector (\ref{150}) is,
in matrix notation (\ref{099}), given by \cite{030}
\be160
{\cal P}_{I ij} =  {\rm e}^{2\phi}\, \alpha_I \, \sigma^1_{ij} \;\;\;,\;\;\;
\sigma^1_{ij} =  
\pmatrix{
 0 & 1  
\cr\noalign{\vskip2mm} 
 1 & 0 }_{ij} \;\;,
\ee
where 
the dilaton is given by ${\rm e}^{-2\phi} = S + \bar S
-2 (C + \bar C)^2$. 

It can be shown \cite{210} that the gauging of a $U(1)$ subgroup of the
$SU(2)$ R-symmetry results in a Killing prepotential of the form
${\cal P}_I =  \alpha_I \, \sigma^1$, where the $\alpha_I$ are again
constant and real.  Thus, the Abelian gaugings we will consider in this
paper are characterised by a Killing prepotential of the form
\be165
{\cal P}_{Iij} =  {\rm e}^{s \phi}\, \alpha_I \, \sigma^1_{ij} \;\;\;,\;\;\;
s= 0, 2 \;\;.
\ee

The class of spacetime metrics that 
we will consider in the following is given by
\be100
ds^2 = {\rm e}^{2 U(z)} {\hat g}_{mn} 
dx^m dx^n + {\rm e}^{-2 p U(z)} dz^2 
\ee
with ${\hat g}_{mn} = {\hat g}_{mn} (x^m)$.  Here 
we denote spacetime indices by $x^{\mu} = (x^m, z), x^m = (t,x,y)$,
and the corresponding tangent space indices by $a = (0,1,2,3)$.
The constant $p$ is introduced for later convenience.
We assume Lorentz invariance in the three-dimensional 
subspace $a = (0,1,2)$.  The metric 
${\hat g}_{mn}$ is thus a three-dimensional constant curvature metric.
We take the solutions to be uncharged, that is we set the gauge fields
to zero.  In general, however, as a consequence of the gauging of (some of)
the isometries of ${\cal M}_H$, 
some of the hyper scalar fields are charged and therefore they 
act as sources for the gauge fields.  The associated currents 
are given by $h_{uv} k^u_I (dq^v + k^v_J
A^J)$. Then, in order for a vanishing gauge field to be 
a consistent solution of its equation of motion, either
the associated current has to vanish or
the charged hyper scalars have to satisfy the equation
$h_{uv} k^u_I dq^v =0$. 
This implies that the hyper scalar fields that are not constant
have to be neutral.  This will indeed turn
out to be the case for the gauging 
(\ref{150}), where the only non-trivial hyper scalar field is the
neutral dilaton field $\phi$.

In the absence of gauge fields, the supersymmetry
transformation laws for the gravitini $\psi_{\mu i}$, 
the gaugini $\lambda^A_i$ ($A = 1, \dots, N_V$) and the hyperini 
$\zeta_{\alpha}$ read \cite{250, 210}
\bea
\delta \psi_{\mu i} &=& {\cal D}_{\mu} \epsilon_i - \ft{1}{2}  \, 
{\cal P}_{Iij} \, {\bar L}^I
 \, \gamma_{\mu} \, \epsilon^j \;\;\;, \;\;\;
\nonumber\\
\delta \lambda^A_i &=& \gamma^{\mu} \partial_{\mu} z^A \epsilon_i
+  \, {\cal P}_{Iij} \, g^{A \bar B} \, {\bar D}_{\bar B} {\bar L}^I
\, 
\epsilon^j  \;\;\;,\nonumber\\
\delta \zeta_{\alpha} &=& V_{u \alpha}^i \gamma^{\mu} {\partial}_{\mu} q^u
\varepsilon_{ij} \epsilon^j -  
2  V_{u \alpha}^i \, k^u_I \, L^I \,
\epsilon_i \;\;.
\label{susy}
\eea
Here we have denoted
the scalar fields residing in the abelian vector multiplets by $L^I$.
For later convenience we now
also introduce a non-holomorphic section which we write 
as ${V}^T = (L^I , M_I)$.  It is
subject to the symplectic constraint 
$i ({\bar L}^I M_I - L^I {\bar M}_I ) = 1$.
The assigment of chiral weights $c$ is as follows.  
The non-holomorphic 
section $V$
has $c = -1$, whereas the supersymmetry parameters
$\epsilon_i$ and $\epsilon^i$ have 
$c = \ft{1}{2}$ and $c = - \ft{1}{2}$, respectively.

Inspection of (\ref{susy}) shows that the following quantity
will play a central
role when solving the equations resulting from the vanishing of the
supersymmetry transformation laws (\ref{susy}):
\bea
W = {\rm e}^{s \phi} \, \alpha_I L^I \;\;\;,\;\;\; s = 0, 2 \;\;\;.
\label{125}
\eea

Supersymmetric domain wall solutions will, in general, only preserve
part of ${\cal N}=2$ 
supersymmetry.  A condition on the supersymmetry parameters
that is consistent with local Lorentz invariance in the subspace $a = 
(0,1,2)$ (and, hence, with (\ref{100})) is the following,
\be130
\epsilon_i = A_{ij} \, \gamma_3 \, \epsilon^j \;,
\ee
where $A_{ij}$ denotes an $SU(2)$ matrix.  Since the spacetime metric 
(\ref{100}) only has a non-trivial dependence on the coordinate 
$z$, we have
$A_{ij} = A_{ij}(z)$. 
Taking the complex conjugate of (\ref{130}) and 
recalling that $(\epsilon_i)^{\star} = \epsilon^i$
(as well as using a convention where $\gamma_3$ is
real) then yields that 
$(A^* A)^i_{j} \, \epsilon^j = \epsilon^i$.  Let us now assume
that (\ref{130}) is the only restriction on the  
supersymmetry parameters, so that $(A^* A)^i_{j} = \delta^i_j$.
On the other hand, 
inserting (\ref{130}) into
the gravitino variation $\delta \psi_{m i}$ and demanding its vanishing 
yields $A_{ij} \sim {\cal P}_{I ij}{\bar L}^I$.
Now, since for general 
gaugings the  condition $(A^* A)^i_{j} = \delta^i_j$ is in general 
not satisfied,
we conclude that for general gaugings 
(\ref{130}) cannot be the only restriction on the supersymmetry
parameters, but that there are further constraints
leading to an additional
(or possibly complete) 
breaking of supersymmetry.
For the Abelian gaugings (\ref{165}) 
considered in this paper, however,  $(A^* A)^i_{j} = \delta^i_j$ is
satisfied, so that 
(\ref{130}) is the only condition on the supersymmetry parameters, thus
leading to
domain wall solutions which preserve $1/2$ of ${\cal N}=2$ supersymmetry.


\section{Solving the BPS equations}
\setcounter{equation}{0}

In this section we will construct supersymmetric 
domain wall solutions (\ref{100}) for
the Abelian gaugings specified by (\ref{150}) and
(\ref{165}).  

The symplectic extension of (\ref{125}) is given by
\be310
W =  
{\rm e}^{s \phi} \, \Big( \alpha_I L^I - \beta^I
M_I \Big) \;\;\;,\;\;\; s = 0, 2 \;\;\;.
\ee 
Here the $\beta^I$ denote real constants associated with the
dual Killing vector $k^I$ and 
the dual Killing 
prepotential ${\cal P}^I_{ij}$.
The quantity (\ref{310}) 
is symplectically invariant, provided that 
$(\alpha_I, \beta^I)$ 
transforms as a vector under symplectic transformations.
In the context of string theory, a constant vector $(\alpha_I, \beta^I)$ 
does
indeed
arise when turning on $H$-fluxes on Calabi-Yau threefolds \cite{020,030}.
Moreover, 
in the presence of these fluxes, it is then possible to engineer
supersymmetric 
domain wall solutions by wrapping D-branes around appropriate supersymmetric
cycles.  We take this to be an indication that, in general, supersymmetric
domain wall solutions do depend on both $\alpha_I$ and $\beta^I$.

We will therefore consider the following symplectic extension of 
the supersymmetry transformation rules (\ref{susy}) (which
are valid in the absence of gauge fields), 
\bea
\label{susyvar}
\delta \psi_{\mu i} &=& {\cal D}_{\mu} \epsilon_i - \ft{1}{2} \, {\bar W} 
\, \sigma^1_{ij} \, \gamma_{\mu} \, \epsilon^j \;\;\;, \;\;\;
\nonumber\\
\delta \lambda^A_i &=& \gamma^{\mu} \partial_{\mu} z^A \epsilon_i
+  
g^{A \bar B} 
\,   {\bar D}_{{\bar B}} {\bar W}
\, \sigma^1_{ij} \,
\epsilon^j  \;\;\;,
\nonumber\\
\delta \zeta_{\alpha} &=& V_{u \alpha}^i \Big(\gamma^{\mu} \partial_{\mu} q^u
\varepsilon_{ij} \epsilon^j - 2 (k_I^u L^I - k^{Iu} M_I) 
\epsilon_i \Big) \;\;,
\eea
with $W$ given by (\ref{310}).
The BPS solutions we will construct in this section are obtained by demanding
the vanishing of the supersymmetry variations (\ref{susyvar}), subject
to a condition of the form (\ref{130}).
It should be pointed out, however, 
that we are not aware of any construction of gauged supergravity
theories giving rise to supersymmetry transformation rules of the type
(\ref{susyvar}) involving
not only ${\cal P}_I$ and $k_I^u$ but also their duals.  Our justification for
starting with (\ref{310}) and (\ref{susyvar}) is twofold.
On the one hand we will be able to solve the system
of equations resulting from the vanishing of (\ref{susyvar}).  On the other 
hand, as stated above, there are examples of 
supersymmetric 
domain wall solutions in string theory which 
are characterised by both the parameters $\alpha_I$ and $\beta^I$ appearing
in (\ref{310}), and we would like to able to reproduce them.

We now impose the following condition on the supersymmetry parameters,
in accordance with the discussion given below (\ref{130}),
\bea
\epsilon_i = {\bar h} \, \sigma^1_{ij} \, \gamma^3 \, \epsilon^j \;\;,
\label{susycond}
\eea
where $h=h(z)$ denotes a phase factor with chiral weight $c= -1$.

Let us first consider the variation $\delta \psi_{m i} =0$.
For a spacetime metric of the form (\ref{100}), we find that 
\bea
\omega_m\,^{ab} = {\hat \omega}_m\,^{ab} + 
{\rm e}^{(p +1) U} \, \partial_z U \, 
\Big (\eta^a_3 \eta^b_m
- (a \leftrightarrow b ) \Big) \;\;\;,
\eea
where ${\hat \omega}_m\,^{ab}$ denotes the spin connection associated to
the three-dimensional metric
${\hat g}_{mn} = {\hat e}_m\,^a {\hat e}_m\,^b \eta_{ab}$, where
$a = 0,1,2$.  The metric ${\hat g}_{mn}$ is a constant curvature metric.
In the anti-de-Sitter case, we may write the constant curvature
condition as
\be300
{\hat R}_{mn}\,^{ab} = \ft{4}{l^2} \Big( {\hat e}_m\,^a {\hat e}_n\,^b
- ( m \leftrightarrow n) \Big) \;,
\ee
where $l$ denotes a real constant, which is related
to the three-dimensional cosmological constant $\Lambda_3$
by $\Lambda_3 = l^{-2}$.  The associated curvature scalar is then 
${\hat R} = 24 \, 
\Lambda_3$.
The condition (\ref{300}) can be viewed \cite{Coussaert:1994jp}
as the integrability condition associated with
\bea
{\hat {\cal D}}_m \, ( h^{1/2} \, \epsilon_i ) = \ft{i}{l} \, 
{\hat e}_m\,^a \gamma_a \gamma_3  \, (h^{1/2} \,\epsilon_i) \;.
\label{btzkill}
\eea
The case of zero curvature is obtained from the above by sending
$l \rightarrow \infty$, whereas the de-Sitter 
case is obtained by replacing $ l \rightarrow i l $.  
The latter does not lead to supersymmetric domain wall solutions.
We also note that the imposition of (\ref{btzkill}) does not lead to
a further restriction of the residual supersymmetry preserved by
(\ref{100}).

Using (\ref{susycond}) as well as (\ref{btzkill}) we then obtain from 
$\delta \psi_{m i} =0$,
\bea
 h {\bar W} = 
{\rm e}^{p U} \partial_z U + \ft{2i}{l} \, {\rm e}^{-U} \;\;.
\label{uw}
\eea
Next, let us consider the variation $\delta \psi_{z i} =0$.
It yields
\bea
{\cal D}_z \epsilon_i = \ft{1}{2} \, {\rm e}^{- p U} \, h {\bar W} 
\, \sigma^1_{ij} \, \gamma^3
\, {\bar h} \, \epsilon^j \;\;.
\label{killing}
\eea
Inserting (\ref{susycond}) on the r.h.s. of 
(\ref{killing}) 
yields
\bea
{\cal D}_z \epsilon_i = \ft{1}{2} \, {\rm e}^{-p U} \, h {\bar W} 
   \, \epsilon_i \;\;.
\label{killing1}
\eea
Hermitian conjugation then gives
\bea
{\cal D}_z \epsilon^i = \ft{1}{2} \, 
{\rm e}^{-p U} \, {\bar h} {W}  \, \epsilon^i \;\;.
\label{killingher}
\eea
On the other hand, inserting (\ref{susycond}) on the l.h.s. of 
(\ref{killing}) yields
\bea
{\cal D}_z \epsilon^i = \ft{1}{2} \, {\rm e}^{- p U} \, h {\bar W}   
\, \epsilon^i - 
(h {\cal D}_z {\bar h}) \, \epsilon^i \;\;.
\label{killing2}
\eea
Comparison of (\ref{killingher}) with (\ref{killing2}) then gives
\bea
h {\cal D}_z {\bar h} = \ft{1}{2} \, {\rm e}^{- p U} \, \Big(
h {\bar W} - {\bar h} W \Big) = \ft{2i}{l} \, {\rm e}^{-( 1 + p) U} \;.
\label{kaehlercon}
\eea
Also, using $\omega_z\,^{ab} =0$ 
and that the pullback of the $SU(2)$ connection is trivial (as we will see 
later),
we find that (\ref{killing1}) is solved by
\bea
h^{1/2} \epsilon_i =  \chi_i \;
{\rm e} ^{\ft{1}{2} U }\;,
\eea
where $\chi_i$ denotes a spinor satisfying (\ref{btzkill}).

Next, let us consider the vanishing of the gaugini variation, 
$\delta \lambda^A_i =0$.
Using the special geometry relations \cite{210}
\bea
g_{A {\bar B}} = - 2 D_A L^I \, {\rm Im} {\cal N}_{IJ} \, 
{\bar D}_{\bar B}
{\bar L}^J \;\;\;,\;\;\;
D_A M_I = {\bar {\cal N}}_{IJ} D_A L^J \;\;\;,\;\;\;
{\bar D}_{\bar A} V =0
\eea
as well as (\ref{susycond}) 
we obtain from $\delta \lambda^A_i =0$,
\bea
g_{A \bar B} \, \partial_z z^A = i \Big[ {\cal D}_z L^I \, {\bar D}_{\bar B} 
{\bar M}_I - {\cal D}_z M_I \, {\bar D}_{\bar B} {\bar L}^I \Big]
 = - {\rm e}^{- p U} \, h \, {\bar D}_{\bar B} {\bar W} \;\;
\label{zgw}
\eea
and, hence,
\bea
\Big( {\bar h} {\cal D}_z L^I + i { \beta}^I \, 
{\rm e}^{s \phi -p U} \Big) \,
{\bar D}_{\bar B} {\bar M}_I = 
\Big( {\bar h} {\cal D}_z M_I + i  { \alpha}_I 
\, {\rm e}^{s \phi -p U} \Big) \,
{\bar D}_{\bar B} {\bar L}^I \;\;.
\eea
Using ${\bar D}_{\bar B} ({\bar L}^I M_I - L^I {\bar M}_I ) = 0$ we may
rewrite this as
\bea
\Big( {\partial}_z ({\bar h} L^I) + i  { \beta}^I 
\, {\rm e}^{s \phi - p U} \Big) \,
{\bar D}_{\bar B} {\bar M}_I = 
\Big({\partial}_z ({\bar h} M_I) + i  { \alpha}_I 
\, {\rm e}^{s \phi - p U} \Big) \,
{\bar D}_{\bar B} {\bar L}^I \;\;,
\eea
or equivalently  as 
$\Pi^T \, \Omega \, {\bar U}_{\bar B} =0$, where $\Omega$ denotes the sympletic
metric and where $\Pi$ and ${\bar U}_{\bar B}$ denote two symplectic vectors,
as follows,
\be330
\Pi = 
   \pmatrix{{\partial}_z ({\bar h} L^I) 
+ i  {\beta}^I \, {\rm e}^{s \phi - p U}
\cr\noalign{\vskip2mm} 
{\partial}_z ({\bar h} M_I) 
+ i  {\alpha}_I \, {\rm e}^{s \phi - p U}} 
\;\;\;,\;\;\;
 {\bar U}_{\bar B} = 
\pmatrix{ {\bar D}_{\bar B} {\bar L}^I 
\cr\noalign{\vskip2mm} 
{\bar D}_{\bar B} {\bar M}_I 
} 
\;\;\;,\;\;\; \Omega = \pmatrix{ 0 & - {\bf 1}
\cr\noalign{\vskip2mm} {\bf 1} & 0 } \;\;\;.
\ee
We note that $\Pi$ has chiral weight $c=0$.

The general solution to the equation $\Pi^T \, \Omega \, {\bar U}_{\bar B} =0$
constructed out of the symplectic vectors $V, {\bar V}, U_A$ and ${\bar U}_A$
is given by
\be340
\Pi = C\,  {\bar h} V + D \, h {\bar V} + E \, h 
(\partial_z {\bar z}^{\bar A}) \, {\bar U}_{\bar A} \;\;\;,
\ee
with  parameters $C, D$ and $E$.
This is so because
$V^T \, \Omega \, {\bar U}_{\bar B} = 0 $ as well as
$ {\bar V}^T \, \Omega \, {\bar U}_{\bar B} = 0 $, whereas
$U^T_A \, \Omega \, {\bar U}_{\bar B} = i g_{A \bar B} \neq 0$. 
We now determine
$C$ and $D$ by considering the symplectic product of (\ref{340}) 
with $h {\bar V} $ and 
${\bar h} V $, i.e. $h {\bar V}^T \, \Omega \, \Pi$ and
${\bar h} V^T \, \Omega \, \Pi$.
We then find that 
\be360
C = - \partial_z U  \;\;\;,\;\;\; D = \partial_z U - h {\cal D}_z {\bar h}
\ee  
by virtue of 
(\ref{uw}) and (\ref{kaehlercon}).

The parameter $E$, on the other hand, can be
determined by contracting $\Pi$ with ${\bar h} U_A$, 
i.e. from ${\bar h} U^T_A \, \Omega
\, \Pi$.  Using $U_A^T \, \Omega \,
{\bar U}_{\bar B} =i g_{A {\bar B}}$ 
as well as
(\ref{zgw}) then yields 
\be370
E=1 \;\;.
\ee
Thus, it follows from (\ref{340}) that
\bea
   \pmatrix{{\partial}_z ({\bar h} L^I )
+ 
\partial_z U \, {\bar h} L^I 
\cr\noalign{\vskip2mm}  
{\partial}_z ({\bar h} M_I ) 
+
 \partial_z U \, {\bar h} M_I }  
= &-& i \, {\rm e}^{s \phi -p U} \, 
\pmatrix{
 { \beta}^I 
\cr\noalign{\vskip2mm} 
 { \alpha}_I }
+ h \partial_z {\bar z}^{\bar B} 
\pmatrix{ {\bar D}_{\bar B} {\bar L}^I 
\cr\noalign{\vskip2mm} 
{\bar D}_{\bar B} {\bar M}_I 
}
\nonumber\\
&+&  \partial_z U \, 
\pmatrix{
h {\bar L}^I 
\cr\noalign{\vskip2mm} 
h {\bar M}_I  } 
-  h {\cal D}_z {\bar h}
\pmatrix{{h} {\bar L}^I 
\cr\noalign{\vskip2mm} 
{h} {\bar M}_I } \;.
\eea
This we rewrite as
\bea
\pmatrix{{\partial}_z Y^I 
\cr\noalign{\vskip2mm} 
\partial_z F_I  }
= &-& i \,  {\rm e}^{s \phi + (1-p)  U} \,
\pmatrix{
  { \beta}^I 
\cr\noalign{\vskip2mm} 
 { \alpha}_I } 
+
{\rm e}^{U} h \partial_z {\bar z}^{\bar B} 
\pmatrix{ {\bar D}_{\bar B} {\bar L}^I 
\cr\noalign{\vskip2mm} 
{\bar D}_{\bar B} {\bar M}_I 
} 
\nonumber\\
&+& \partial_z U \, 
\pmatrix{
{\bar Y}^I 
\cr\noalign{\vskip2mm} 
{\bar F}_I  }  -
h {\cal D}_z {\bar h}
\pmatrix{{\bar Y}^I 
\cr\noalign{\vskip2mm} 
{\bar F}_I } \;\;,
\label{stab1}
\eea
where we introduced the chiral 
invariant variables 
\bea
Y^I = {\rm e}^U \, {\bar h} \, L^I \;\;\;,\;\;\; 
F_I(Y) = {\rm e}^U \, {\bar h} \,
M_I \;\;\;.
\label{yf}
\eea
Inserting (\ref{yf}) into
$i ({\bar L}^I M_I - L^I {\bar M}_I ) = 1$ yields
\bea
{\rm e}^{2U} = i \Big[{\bar Y}^I F_I(Y)  
- Y^I {\bar F}_I ({\bar Y}) \Big]  \;\;.
\label{u}
\eea
Using
\bea
{\rm e}^{U} h \partial_z {\bar z}^{\bar B} 
\pmatrix{ {\bar D}_{\bar B} {\bar L}^I 
\cr\noalign{\vskip2mm} 
{\bar D}_{\bar B} {\bar M}_I } 
=
{\rm e}^{U} h {\cal D}_z 
\pmatrix{ {\bar L}^I 
\cr\noalign{\vskip2mm} {\bar M}_I } 
= 
\partial_z 
\pmatrix{
{\bar Y}^I 
\cr\noalign{\vskip2mm} 
{\bar F}_I  } - \partial_z U 
\pmatrix{
{\bar Y}^I 
\cr\noalign{\vskip2mm} 
{\bar F}_I  } - {\bar h} {\cal D}_z h   \pmatrix{
{\bar Y}^I 
\cr\noalign{\vskip2mm} 
{\bar F}_I  } 
\eea
it follows from (\ref{stab1}) that
\bea
\partial_z
\pmatrix{Y^I - {\bar Y}^I
\cr\noalign{\vskip2mm} 
F_I  - {\bar F}_I}
=  - i  \, {\rm e}^{s \phi + (1-p)  U} \,
\pmatrix{
 { \beta}^I 
\cr\noalign{\vskip2mm} 
{\alpha}_I } \;.
\label{ypar}
\eea
If we now set 
\be400
 {\rm e}^{s \phi} = {\rm e}^{(p-1)  U} \;\;,
\ee
then we may integrate (\ref{ypar}),
\bea
\pmatrix{Y^I - {\bar Y}^I
\cr\noalign{\vskip2mm} 
F_I - {\bar F}_I }
= i 
\pmatrix{
h^I -  \beta^I z
\cr\noalign{\vskip2mm} 
h_I - \alpha_I z }
= i \pmatrix{H^I
\cr\noalign{\vskip2mm} 
H_I }  \;\;,
\label{stab}
\eea
where the $(H^I, H_I)$ denote harmonic functions.
These equations, which determine the behaviour of the physical vector scalar
fields $z^A =Y^A/Y^0$, are also known from the study of BPS black hole 
solutions in ungauged
${\cal N}=2$ supergravity theories, where they are called attractor equations
\cite{Fer:1996dd,Fer:1996um}.

Let us now return to (\ref{400}).  For the case $s=0$
(corresponding to an Abelian gauging of the $SU(2)$ R-symmetry), (\ref{400})
holds for the convenient choice $p =1$.  For the case $s=2$ (corresponding
to the Abelian gauging of a particular isometry of the universal hyper moduli
space)
we will show
below that (\ref{400}) solves the equation resulting from the vanishing of
the hyperino variation, provided that $p=-3$.

Next, consider inserting (\ref{stab}) into (\ref{kaehlercon}).  This yields 
\bea
l^{-1} = - \ft{1}{4} \, {\rm e}^{s \phi} \, ( \alpha_I h^I - \beta^I h_I) \;\;.
\label{cosm}
\eea
Since $l$ has to be constant, we conclude that 
it is not possible to have a non-vanishing three-dimensional cosmological
constant $\Lambda_3$ for the case $s=2$.  That is, for $s=2$ the domain wall
solutions 
have to be flat ($ \alpha_I h^I - \beta^I h_I =0$), whereas for $s=0$
they may be curved.  This implies that in the latter case there is no 
restriction on the integration constants $(h^I, h_I)$, whereas in the
former case there are such restrictions.

Before turning to the hyperini variation $\delta \zeta_{\alpha}=0$, 
let us perform two consistency checks on the solution given above.
Let us first return to (\ref{uw}).  Using (\ref{ypar}) as well as (\ref{400})
the real part of (\ref{uw}), $h {\bar W} + {\bar h} W = 2
{\rm e}^{p U} \partial_z U$, may be rewritten as 
\bea
\partial_z {\rm e}^{2U} =  \Big( { \alpha}_I (Y^I + {\bar Y}^I)
- { \beta}^I (F_I + {\bar F}_I) \Big) = i \partial_z
\Big({\bar Y}^I F_I - Y^I {\bar F}_I \Big) \;\;,
\eea 
which is in perfect agreement with (\ref{u}). Next, let us consider
$h {\cal D}_z {\bar h} = h \partial_z {\bar h} - i {\cal A}_z$, where
${\cal A}_z = {\cal A}_z^Y - \ft{i}{2} \partial_z \log \ft{\bar h}{h} \,,\, 
{\cal A}_z^Y =  \ft{1}{2} {\rm e}^{-2U} (Y^I - {\bar Y}^I) 
\stackrel{\leftrightarrow}
\partial_z
(F_I - {\bar F}_I )$.  Using (\ref{stab}) it  follows that 
$h {\cal D}_z {\bar h} = -i {\cal A}_z^Y = - \ft{i}{2}  {\rm e}^{-2U} 
( { \alpha}_I h^I - { \beta}^I h_I)$ which, upon using  
(\ref{cosm}), precisely yields (\ref{kaehlercon}).

Finally , let us turn to the variation 
$\delta \zeta_{\alpha}=0$.  For the case $s=0$ we have $k_I^u = 0, k^{Iu} =0$,
so that $\delta \zeta_{\alpha}=0$ is solved by constant hyper scalars,
$q^u = {\rm const}$.  On the other hand, for $s=2$ we have 
$k_I^{S - {\bar S}} = - 2 \,
\alpha^I$ and $ k^{I \, S - {\bar S}} = - 2 \beta^I$.
We solve the hyperino variation by setting $ S - {\bar S}
= {\rm const}$ and $C = {\rm const}$, so that also in this case 
the pullback of the $SU(2)$
connection is trivial.  Then, using
$\partial_{\mu} (S + {\bar S}) 
= \partial_{\mu}
{\rm e}^{-2\phi}$, it follows that
\bea
\delta \zeta_{\alpha} &=& V_{S + {\bar S} \,,\alpha}^i \, 
\gamma^{\mu} \partial_{\mu} {\rm e}^{-2\phi} \,
\varepsilon_{ij} \, \epsilon^j + 4  \, V_{S - {\bar S} \,,\alpha}^i \, 
({\alpha}_I L^I 
- {\beta}^I M_I) \,
\epsilon_i \,.
\eea
Using
\bea
V_{S + {\bar S}}^{i \alpha} = \frac{1}{2 \sqrt{2}}
\pmatrix{0 & {\rm e}^{2 \phi} 
\cr\noalign{\vskip2mm} 
 {\rm e}^{2 \phi} & 0 }^{i \alpha} \;\;\;,\;\;\;
V_{S - {\bar S}}^{i \alpha} = \frac{1}{2 \sqrt{2}}
\pmatrix{0 & - {\rm e}^{2 \phi} 
\cr\noalign{\vskip2mm} 
 {\rm e}^{2 \phi} & 0 }^{i \alpha} \;\;
\eea
as well as (\ref{susycond})
and $\varepsilon_{12}=1$, 
we obtain 
\bea
{\rm e}^{p U} \partial_z {\rm e}^{-2\phi} = 4  \, {\bar h} \,
({ \alpha}_I L^I 
- { \beta}^I M_I) = 4 \, {\rm e}^{-2 \phi} \, {\bar h} W\;.
\label{eqphi}
\eea
It follows that ${\bar h} W = h {\bar W}$, and hence we obtain from (\ref{uw})
that $l^{-1} = \infty$.  Using (\ref{uw}) it follows that 
(\ref{eqphi}) is solved by
\bea
{\rm e}^{-2 (\phi - \phi_0)} = {\rm e}^{4U} \;.
\label{dila}
\eea
This is consistent with (\ref{400}) provided that $p =-3$.

Let us briefly summarize some of the properties of 
the domain wall solutions we have constructed.
For both cases ($s = 0,2$) we find that the behaviour of the vector scalar
fields is determined by a set of attractor equations given in (\ref{stab}).
For the case $s=0$, the hyper scalars are all constant and the domain wall
solutions are, in general, curved, with the three-dimensional cosmological
constant $\Lambda_3 = l^{-2}$ determined by (\ref{cosm}).  The associated
spacetime metric is given by (\ref{100}) with $p = 1$.  For the case $s=2$,
on the other hand, we find that the dilaton field exhibits the run-away 
behaviour (\ref{dila}).  The domain wall solutions must be flat, and the
associated 
spacetime metric is given by (\ref{100}) with $p = -3$.


\section{Discussion}

\setcounter{equation}{0}


The domain wall solutions specified by equations (\ref{100}), 
(\ref{u}), (\ref{stab}) and (\ref{cosm}) solve
the Killing spinor equations for any prepotential 
function describing the couplings of abelian vector multiplets to
supergravity.  In the
case that only the R-symmetry is gauged ($s=0$), the hyper scalars are
constant
and the solutions possess an asymptotic ($z \rightarrow \infty$) AdS vacuum. 
In the case $s=2$ (corresponding to the gauging of an axionic shift symmetry),
on the other hand, the dilaton is not constant but instead given by 
(\ref{dila}).  Thus, it is not stabilized by the potential.
In either
case, the behaviour of the 
scalars in the vector multiplets is determined by the same set of attractor
equations (\ref{stab}) known from the study of 
BPS black holes in ungauged supergravity. 
Moreover, the 
domain wall may possess constant AdS-like curvature according to 
(\ref{cosm}).  This feature also has a counterpart in single-center
black hole solutions, namely in the angular momentum carried by
these stationary solutions. 
There is the following one-to-one 
correspondence between quantities determining 
black hole solutions and domain
wall solutions with non-trivial vector scalars:

\begin{center}
\begin{tabular}{lc|cl}
black hole                  &\quad & \quad & domain wall \\ \hline
central charge $Z$          &&& superpotential $W$ \\
entropy   ${\cal S}$               &&& cosmological constant $V_{extr.}$ \\
angular momentum  $J$    &&   & constant wall curvature $\Lambda_3$
\end{tabular}
\end{center}

This analogy is related to the fact that both types of solutions can be
reduced to equivalent one-dimensional 
systems \cite{230,190}.

The fact that all the 
regular critical points are reached at the boundary of the
AdS space (${\rm e}^{2U} \rightarrow \infty$) excludes the possibility of 
a regular RG-flow
and/or of trapping of gravity near the wall \cite{240,280}.  {From} the
RG-flow point of view the AdS vacuum corresponds to an UV fixed point,
and when moving towards negative $z$ the warp factor ${\rm e}^{2U}$ decreases
monotonically (in accordance with a $c$-theorem). If we do not add a source,
then ${\rm e}^{2U}$ will have to pass through a zero 
at some finite value of $z$, which describes a 
singular end-point of the RG flow.  As in any other cases with
vector multiplets only, the absence of an IR fixed point forces the
solution to run into a singularity.

On the other hand, in the context of string theory,
it is possible to engineer domain wall solutions with a 
non-trivial dilaton field in terms of D-branes wrapped around 
supersymmetric cycles.  It is thus reasonable to put
appropriate sources at some place where the warp factor is still
positive, say at $z=0$.  These source terms will then appear on the 
right-hand side of the harmonic equations
\be184
\partial^2 H_I \sim \alpha_I \, \delta(z) \quad , \quad 
\partial^2 H^I \sim \beta^I \, \delta(z) \;\;,
\ee
where ($\alpha_I, \beta^I$) are 
the charges in a basis of the wrapped cycles. 
For dilatonic walls that are flat (which correspond to the case $s=2$
considered in this paper), there
are two ways in which one can solve (\ref{184}).
The first possibility consists in continuing the solution through the source 
in a symmetric way, which
implies the replacement
%
$H_I \rightarrow h_I + \alpha_I \, |z|\,,\, 
H^I \rightarrow h^I + \beta^I \, |z|$,
%
and which is equivalent to a sign change in the flux vector ($\alpha_I,
\beta^I$) while passing through the source at $z=0$.  In this case the scalar
fields and the warp factor ${\rm e}^{2U}$, which are 
determined from (\ref{stab}), are $Z_2$-symmetric.
One may, however, also consider the case where the flux jumps from zero to a
finite value, i.e.\ on the side behind the source we can set
$\alpha_I=0, \beta^I =0$, which is equivalent to performing the replacement
$H_I \rightarrow h_I + \alpha_I \, {1 \over2} (z + |z|)$, $H^I
\rightarrow h^I + \beta^I \, {1 \over2} (z + |z|)$, so that 
for negative $z$ the $H_I$ and 
$H^I$ are constant and the spacetime is flat.  
Thus, by adding sources, we cut off the (singular) part
of the spacetime and instead glue on either 
an identical piece (yielding a $Z_2$ symmetric solution)
or flat spacetime (corresponding to vanishing flux on one side of the wall).

Let us discuss a few concrete examples of dilatonic domain walls.
Consider the case 
where the only non-vanishing components of $(\alpha_I, \beta^I)$
are given by $\alpha_0$ and
$\beta^A$ ($A = 1,2,3$).  In 
type IIA theory, the corresponding
domain wall solution can then be obtained by a torus
compactification of the D-brane intersection $2\times 6 \times 6
\times 6$, where the 6-branes wrap a 4-cycle in the internal space and where
the common 2-brane represents the domain wall in the four non-compact
directions. The dual configuration with non-vanishing $\alpha_A$ ($A=1,2,3$) 
and $\beta^0$ is described by the 
intersection $4 \times 4 \times 4 \times 8$, where the 4-branes wrap
a 2-cycle and where the 8-brane wraps the whole internal space. These solutions
can of course also be mapped onto the type IIB side, where
they are given by the  
brane intersections $5\times 5 \times 5 \times 5$ and $3
\times 5 \times 5 \times 7$, respectively. This picture in terms of brane
intersections is the one appropriate for torus compactifications.  For
generic Calabi-Yau threefold 
compactifications, a domain wall solution with non-vanishing
$\alpha_0$ and $\beta^A$ is obtained on the type IIA side 
by wrapping a single 2- and a single 
6-brane around a 
holomorphic 0- and a holomorphic 4-cycle of the Calabi-Yau manifold.
Similarly, on the type IIB side, a domain wall solution is obtained
by wrapping a single 5-brane
around a supersymmetric 3-cycle. In either case the symplectic
flux vector ($\alpha_I , \beta^I$) is the decomposition of the corresponding
brane charges in a basis of 2-, 3- or 4-forms.

Let us describe the type IIA solution with non-vanishing 
$\alpha_0$ and $\beta^A$ in some more detail.
In the limit where the volume of the Calabi-Yau threefold is taken to be
large, the associated homogenous function $F(Y)$ is given by
\bea
F(Y) = \frac{D_{ABC} \, Y^A Y^B Y^C }{ Y^0} \;\;\;,\;\;\;
D_{ABC} = - \ft{1}{6}\, C_{ABC} \;\;, 
\eea
where the coefficients
$C_{ABC}$ denote the intersection numbers of the 4-cycles of the threefold
($A = 1, \dots, N_V$).  Solving 
the attractor equations (\ref{stab}) 
in terms of the harmonic functions $H_0$ and $H^A$, and using (\ref{u}), yields
\bea
Y^0 = {\bar Y}^0 \;\;\;,\;\;\; 
 (Y^0)^2 = \frac{D_{ABC} H^A H^B H^C}{ 4 H_0} \;\;\;,\;\;\;
Y^A = \ft{i}{2} H^A\;\;,\;\;\;{\rm e}^{2U} = -4 Y^0 H_0 \;\;\;.
\eea
Since the curvature on the wall (\ref{cosm})
is zero, it follows from (\ref{uw}) that $h {\bar W} = {\bar h} W$.  Using
(\ref{yf}) as well as (\ref{dila}), we then obtain
${\bar h} W = \pm |W| = {\rm e}^{- 5 U} \, (\alpha_0 Y^0 - \beta^A F_A(Y))$,
and hence 
\bea
|W|^2 = \ft{1}{16} \, {\rm e}^{- 10 U} \, |Y^0|^{-2}
|D_{ABC}\, ( \frac{\alpha_0}{H_0} \, H^A H^B H^C  + 3  \beta^A H^B H^C)|^2
\;\;\;.
\eea

There are some issues that we didn't address in this paper
and which may be worth studying in the future.
First, when evaluating 
the pullback of the $SU(2)$ connection 
of the quaternionic space for the $s=2$ solution
we find that it is
trivial. In general, however, we expect that 
it will also contribute to the curvature of
the domain wall. This effect, by the way, may also happen for BPS walls in
five-dimensional  
gauged supergravity. Second, the geometry of the curved wall is
$AdS_3$ and it would be interesting to understand whether a global
identification can give rise to a BTZ black hole.  Third, in the case of 
dilatonic
walls, the curvature on the wall had to be zero 
($\alpha_I h^I - \beta^I h_I = 0$).  It may, however, happen that this
restriction on the curvature is circumvented 
by allowing the dilaton also to depend 
on the worldvolume coordinates.
And finally, it would be interesting to investigate
more general potentials for the dilaton field.  These can
be obtained either by gauging some of the non-compact isometries
of the universal hypermultiplet moduli space or by
turning on some of the NS 3-form fluxes in type IIB
Calabi-Yau compactifications. In particular, $SL(2,{\bf Z})$
covariant superpotentials may eliminate 
the run-away behavior of the dilaton field and hence provide
a vacuum which is stable with respect to the dilaton field.

\bigskip

{\bf Acknowledgements}

We would like to thank G. Curio, G. Dall'Agata and
B. de Wit 
for fruitful discussions.
The work of K.B. is supported by a Heisenberg grant of the DFG.
This work is also supported by the European Commission RTN Programme
HPRN-CT-2000-00131.





\providecommand{\href}[2]{#2}\begingroup\raggedright\endgroup

\end{document}